\documentclass[aps,twocolumn,groupedaddress,prb]{revtex4}

\usepackage{graphicx}

\begin{document}

\author{P. Fraundorf}
\email{pfraundorf@umsl.edu}
\affiliation{Department of Physics and Astronomy and Center for Molecular 
Electronics, University of Missouri - StL, St. Louis, Missouri 63121}

\title{Friendly units for coldness}


\begin{abstract}
Measures of temperature that center around
human experience get lots of use.  Of course thermal physics insights 
of the last century have shown that reciprocal temperature (1/kT) 
has applications that temperature addresses less well. 
In addition to taking on negative absolute values under 
population inversion (e.g. of magnetic spins), bits and 
bytes turn 1/kT into an informatic measure of the thermal 
ambient for developing correlations within any complex system.  
We show here that, in the human-friendly units of bytes and 
food Calories, water freezes when 
1/kT $\cong$ 200 ZB/Cal or kT $\cong$ 5 Cal/YB.  Casting 
familiar benchmarks into these terms shows that habitable human 
space requires coldness values (part of the time, at least) 
between 0 and 40 ZB/Cal with respect body temperature 
$\sim100^\circ$F, a range in kT of $\sim1$ Cal/YB.  
Insight into these physical quantities 
underlying thermal equilibration may prove useful for
budding scientists, as well as the general public, in years 
ahead.
\end{abstract}

\maketitle


\section{Introduction}
\label{sec:Intro}

Bruno Maddox's article \citep{Maddox06} in the June 2006 Discover Magazine on the Nightmare of Divided Loyalties between Fahrenheit and Centigrade is timely for this age of information. That's because thermodynamics texts \citep{Tribus61, Castle65, Girifalco73p, Kittel80p, Stowe84p, Plischke89c, Betts92, Moore98, Schroeder00} (following pre-1960 papers \citep{ShannonWeaver49, Jaynes57a, Jaynes57b} by Shannon and Jaynes have increasingly exploited reciprocal temperature (Garrod's coldness \citep{Garrod95}) to capture the essence of temperature as equilibrant more generally (e.g. for spin systems), and as energy's uncertainty slope i.e. a rate of increase in state uncertainty (information units) per unit increase in thermal energy.  

One reason is that thermal physics insights of the last century \citep{pf.hcapbit} have shown reciprocal temperature (1/kT) to have applications that temperature addresses less well. In addition to taking on negative absolute values under population inversion (e.g. of magnetic spins), bits and bytes turn it into an informatic measure of the thermal ambient (to wit: the mutual information value of free energy)\citep{Tribus71, Lloyd89b, Schneider91b, Bennett2003} for 
developing correlations in all sorts of complex systems (e.g. within 
molecules, galaxies, and biological communities)\citep{Chaisson04}.

\section{Say what?}
\label{sec:howitworks}

How does coldness (1/kT, or energy's Lagrange multiplier) work?  Basically heat flows spontaneously (by chance maximizing uncertainty) from low to high uncertainty slopes that range between minus and plus infinity, at which end points reciprocal coldness (kT) asymptotically goes to absolute zero. Thus one thing traditional treatments miss is that negative absolute temperatures are routinely employed (say in the LASER on your DVD reader, or in a domino toppling contest) by moving up the temperature scale, rather than down through zero. Unlikely inversions (negative uncertainty slopes) even add something to the spectator value of gambling. 

Taken in reverse, it also means that new correlations between subsystems extract a price in the thermalization of available work. The price decreases if the heat created can be dumped into a colder reservoir, making this observation relevant to molecular engines as well as to humans faced with a warming environment.

\section{Calories and bytes}
\label{units}

If (inspired by Bruno's request for something everyone can relate to) you put coldness (1/kT) into everyday units (Calories and bytes) then you'll find that water freezes at almost precisely 200 zettabytes per Calorie where zetta is the SI prefix for $10^{21}$. In other words, resetting a 200 gigabyte DNA string or computer memory to specified values will convert at least a picoCalorie of available work into ice-melting heat. The same operation may generate less heat, if the heat can be dumped at lower temperature. Hence the bit depth of a digital camera can be made larger if its CCD is cooled, and when it's cold outside you in principle really can get more work done.

\begin{figure}[tbp]
\includegraphics[scale=.7]{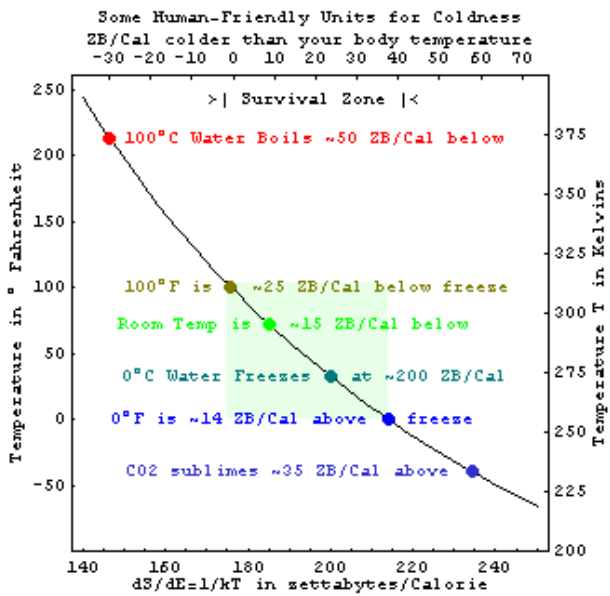}%
\caption{In ZB/Cal, $0^\circ$C is 200.208 with 
$0^\circ$F up by $\sim$14, room temperature down by 
$\sim$15 and $100^\circ$F down by $\sim$25, 
giving us a $\sim40$ ZB/Cal habitable range 
upward from 1/kT internal.}
\label{Fig01}
\end{figure}

As a result one can say (cf. Figure \ref{Fig01}) that ``chilly Europe" at 
$0^\circ$F is about 14 ZB/Cal colder than ice water, which at 
$0^\circ$C is about 15 ZB/Cal colder than room temperature. In turn 
room temperature is about 10 ZB/Cal colder than ``hot Europe" at 
$100^\circ$F, which is approximately the temperature inside your body.  
At ambient temperatures below these, our food-powered metabolism plus some blankets might keep you warm for a while. At ambient temperatures above these our metabolism, rather than keeping us warm, generates waste heat that we must actively export to survive. Food is less useful then. Thus our ``long term survival zone" extends from about 0 to 40 zettabytes 
per Calorie colder than $98.6^\circ$F. Weather persistently outside this 
range is a fundamental problem for each of us, even though it might not 
seem so if you mainly see persistent heat or cold from within a 
temperature-controlled shell.

\section{Other benchmarks}
Boiling water's coldness is around 50 ZB/Cal below ice water's 200 ZB/Cal absolute. Even lower uncertainty slopes occur in the 9 ZB/Cal of our sun's surface, the 0 ZB/Cal of a spin system with equal up/down populations, and the -7 ZB/Cal of a He-Ne LASER's 99$\%$ excited-Ne inversion. Familiar things with really high coldness 
include dry ice at 35 ZB/Cal above freezing, liquid nitrogen at 
nearly 800 ZB/Cal absolute, liquid helium at over 13000 ZB/Cal, and 
our universe's blackbody ambient at around 20000 ZB/Cal. Thus local 
issues notwithstanding, the larger world around us is pretty cold! 
The strange behaviors of liquid nitrogen and helium also reflect 
high uncertainty slopes, and thus the extreme increases in state 
uncertainty that result from adding small amounts of heat.

\section{So what?}

In context of the Discovery article's request for some resolution to the nightmare, energy's uncertainty slope could therefore be the ``hero'' 
that Bruno was looking for \citep{Maddox06} in the face of summer days 
with respect to which our blood {\em must} run cold.  More to the 
point, telling students about physical in addition to historical units
for the parameter that equilibrates on thermal contact could further 
prepare them to recognize other cross-disciplinary connections.  Such 
connections (in cross-cutting fields like 
nanoscience, informatics, and astrobiology) will 
likely prove useful for tackling 
the multiscale challenges that our species 
faces in the days and 
millenia \citep{Ward03} ahead.  

\begin{figure}[tbp]
\includegraphics[scale=.7]{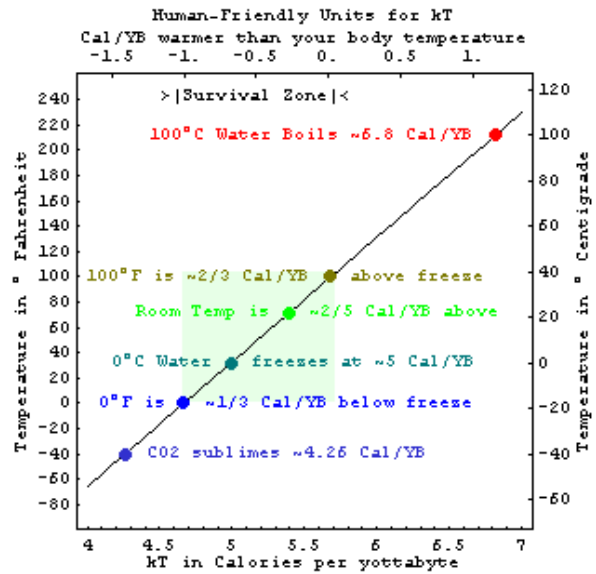}%
\caption{In Cal/YB, $0^\circ$C is 5 with 
$0^\circ$F down by $\frac{1}{3}$, room temperature 
up by $\frac{2}{5}$, and $100^\circ$F up by 
$\frac{2}{3}$ 
giving humans a $\sim1$ Cal/YB habitable range 
downward from our internal kT.}
\label{Fig02}
\end{figure}

Of course given our cultures' tenacity for repetition, plus the fact that kT is a useful measure of energy fluctuation in quadratic systems (e.g. kT is also 1/40 eV/nat at STP), it is unlikely that bytes per Calorie will become a routine part of the morning weather report anytime soon. In that context, perhaps the reader can instead find a mnemonic measure of kT (cf. Fig. \ref{Fig02}) to assist in the clarifying move from historical, to natural, perspectives on that continuum from cool to warm and beyond.

\begin{acknowledgments}
This work has benefited indirectly 
from support by the U.S. Department of Energy, 
the Missouri Research Board, as well as 
Monsanto and MEMC Electronic Materials Companies. 
\end{acknowledgments}

\bibliography{ifzx}






\end{document}